\documentclass[12pt,psf,epsf]{article}
\usepackage[centertags]{amsmath}
\usepackage{amssymb}
\usepackage{graphicx}
\usepackage{epsfig}
\usepackage{ulem}
\usepackage[english]{babel}
\usepackage{array}
\usepackage{amsthm}
\usepackage{latexsym}
\usepackage[mathcal]{euscript}
\pdfoutput=1
\usepackage{epsfig}

 \usepackage{jheppub}
 \usepackage{hyperref}

\usepackage{subfigure}
\usepackage{psfrag}

\usepackage{epsfig}
\usepackage[latin1]{inputenc}
\usepackage{float}
\usepackage{graphicx}
\usepackage{cancel}
\usepackage{mathrsfs}
\usepackage{amssymb}
\usepackage{amsfonts}
\usepackage{amsmath}
\usepackage{slashed}
\usepackage{graphicx}
\usepackage{bm}
\usepackage{color}

\usepackage{setspace}

\def\be{\begin{equation}}
\def\ee{\end{equation}}
\def\bea{\begin{eqnarray}}
\def\eea{\end{eqnarray}}
\def\ba{\begin{array}}
\def\ea{\end{array}}

\def\cQ{\mathbb{Q}}

\DeclareMathOperator{\sign}{sign}

\begin{document}
\title{Coleman-Weinberg potential in $p$-adic field theory}

\author[a]{Dmitry S. Ageev,}
\author[b]{Andrey A. Bagrov,}
\author[c]{Askar A. Iliasov}
\affiliation[a]{Steklov Mathematical Institute, Russian Academy of Sciences, Gubkin str. 8, 119991 Moscow, Russia}
\affiliation[b]{Department of  Physics  and  Astronomy,  Uppsala  University, Box 516,  SE-75120  Uppsala,  Sweden}
\affiliation[c]{Institute for Molecules and Materials, Radboud University, Heyendaalseweg 135, 6525AJ Nijmegen, \mbox{The Netherlands}}

\emailAdd{ageev@mi-ras.ru}
\emailAdd{andrey.bagrov@physics.uu.se}
\emailAdd{a.iliasov@science.ru.nl}

\abstract{In this paper, we study $\lambda \phi^4$ scalar field theory defined on the unramified extension of p-adic numbers ${\mathbb Q}_{p^n}$. For different ``space-time'' dimensions $n$, we compute one-loop quantum corrections to the effective potential. Surprisingly, despite the unusual properties of non-Archimedean geometry, the Coleman-Weinberg potential of p-adic field theory has  structure very similar to that of its real cousin. We also study two formal limits of the effective potential, $p \rightarrow 1$ and $p \rightarrow \infty$. We show that the $p\rightarrow 1$ limit allows to reconstruct the canonical result for real field theory from the p-adic effective potential and provide an explanation of this fact. On the other hand, in the $p\rightarrow\infty$ limit, the theory exhibits very peculiar behavior with emerging logarithmic terms in the effective potential, which has no analogue in real theories.}

\maketitle

\newpage

\section{Introduction}
A possible relevance of non-Archimedean geometry and $p$-adic number theory within different contexts of theoretical physics is being discussed for more than thirty years. Originally p-adic concepts have been introduced in string theory \cite{Volovich:1987wu,Volovich:1987nq,Freund:1987kt,Brekke:1988dg,Frampton:1987sp,Frampton:1988kr} as a model of spacetime beyond the Planck distance, where one should not expect the Archimedean axiom to hold true. The concept of ultrametric spaces and the corresponding mathematical machinery percolated into other fields of knowledge from high-energy \cite{Manin,IA-padic,Arefeva:2004qqr} to condensed matter physics \cite{parisi,kozyrev,Bentsen:2019rlr,Zharkov:2012qea} and biology \cite{bio1, bio2}, since they were appreciated for being a natural language to describe hierarchical systems \cite{VVZ,Dragovich:2017kge}.

Recently, interest in $p$-adic quantum field theory has revived due to the possible relations between geometry of non-Archimedean number fields and the holographic correspondence \cite{Gubser2016}. The $p$-adic version of the $AdS/CFT$ duality with bulk geometry represented by Bruhat-Tits tree has been introduced \cite{Gubser2016,Gubser:2016htz,Heydeman:2016ldy,Stoica:2018zmi}. Since the boundary quantum field theory in this case is defined over a $p$-adic number field (either $\mathbb{Q}_p$ or its unramified extension $\mathbb{Q}_{p^n}$), further development of non-Archimedean holographic duality requires a deeper insight into the structure of $p$-adic field theory. Among other things, the Wilson renormalization group has been studied perturbatively, and critical exponents were computed in hierarchical bosonic and fermionic models \cite{Lerner1,Lerner:1991cd,Lerner:1994th}, for large-N models \cite{GubserON}, and for scalar field theory defined over mixed ($p$-adic/real) number fields \cite{Gubser:2018ath}. In this note, we attempt to make a step further in this direction and construct a non-Archimedean analogue of the Coleman-Weinberg effective potential  \cite{ColemanWeinberg}. The Coleman-Weinberg potential is an important and illustrative concept which allows to incorporate quantum effects on the level of theory action, and provides a natural language to speak of symmetry breaking in interacting field theory \cite{Dolan:1973qd}. The renormalization group flow is convenient to represent in terms of the effective potential as well. In ${\mathbb R}^n$ scalar field theory, one can derive the renormalization group by computing scattering amplitudes or integrating out UV momentum shells in the Wilsonian approach. In $p$-adic field theory, coordinates/momenta and wave functions take values in different number fields, making certain construction normally used to describe RG flows (e.g. the Callan-Symanzik equation) tricky to define. This gives an additional motivation to study the non-Archimedean effective potential.

A critical issue that one almost unavoidably encounters when trying to construct a quantum field theory over $p$-adic numbers is the lack of well-defined space-time signature that makes the very concept of Lorentzian or Euclidean symmetry poorly defined. Pragmatically speaking, it means that Wick rotation cannot be used to bypass difficulties emerging in the Lorentzian case by performing analytical continuation to Euclidean time. For non-Archimedean $AdS_2/CFT_1$ holography, this problem has been addressed in \cite{Stoica-Lorentz}, where a possible approach to defining spacelike and timelike geodesics in the $p$-adic bulk via constructing quadratic extension $\mathbb{Q}_p\left[\sqrt{\tau}\right]$ of the number field, and expanding the original ``spacelike'' Bruhat-Tits tree with a set of branches that are postulated to be timelike. However, to a large extent this problem remains unresolved, especially outside of the holographic context, and one has to cope with it to compute observables in $p$-adic quantum field theory.

The Coleman-Weinberg potential is usually computed in Euclidean signature, while its Lorentzian treatment leads to appearance of certain pathological structures such as logarithmic divergences and imaginary terms in the potential already in field theories defined over ${\mathbb R}^n$. 
Since in the $p$-adic case, there is no way to do analytical continuation, we take the measure of the path integral $e^{-S}$ rather than $e^{iS}$ as a starting point of our consideration.

We compute one-loop effective potential of a real-valued $\lambda\phi^4$ scalar field theory with quadratic dispersion defined on $\mathbb{Q}_{p^n}$ space, mainly focusing on the $n=1$ (``$p$-adic quantum mechanics'' \cite{VVZ}), $n=2$ and $n=4$ cases. The quantum corrections to the potential are given by integrals over ${\mathbb Q}_{p^n}$ that can be expressed as infinite (divergent) series. We find a tractable approximation that allows to evaluate them and, after renormalization, obtain an explicit expression for the effective potential. In all considered dimensions, the resulting potentials have very similar structure to their real analogues. Moreover, in the formal $p\rightarrow 1$ limit, an exact matching occurs. A very peculiar behavior is observed in the opposite, $p \rightarrow\infty$ limit, where the potential acquires logarithmic term $\ln\left(1+\lambda \phi_b^2/2\right)$.
However, given the fact that this takes place in any dimension including $n=1$, which is suspicious even in the rather exotic non-Archimedean setting, we think that this could either be an artifact of the one-loop approximation or should be cured with renormalization group transformation of the potential.

The paper has the following structure. In Sec. \ref{sec:main_section}, we define the model, obtain a formal expression for the Coleman-Weinberg potential, and perform its renormalization for three particular cases, $n=1, \,2,\, 4$. In Sec. \ref{sec:pto1}, we define the formal $p\rightarrow 1$ limit, and use it to relate the p-adic effective potential to its real cousin. In Sec. \ref{sec:ptoinfty}, we consider the $p\rightarrow\infty$ limit. In Sec. \ref{sec:EM_est}, an alternative approach to computing the effective potential via the Euler-Maclaurin formula is proposed, and its validity limits are discussed. Sec. \ref{sec:summary} briefly summarizes the obtained results. App. A contains definition of the unramified extension $\mathbb{Q}_{p^n}$. App. B proves an identity relating integrals over $\mathbb{Q}_{p^n}$ and $\mathbb{R}^n$. App. C is to remind the standard Coleman-Weinberg calculation in real field theory.

\section{Coleman-Weinberg potential in $p$-adic field theory} \label{sec:main_section}
We shall focus on the real-valued scalar field theory defined over the unramified extension $\mathbb{Q}_{p^n}$ of $p$-adic number field:
\begin{equation}\label{eq:main}
    S = \int\limits_{{\mathbb Q}_{p^n}}dk \widetilde{\phi}(-k) ( |k|^{s}) \widetilde{\phi}( k)+\frac{\lambda}{4!} \int\limits_{\mathbb{Q}_{p^n}}dx  \phi(x)^4, \,\,\,\, x \in {\mathbb{Q}}_{p^n}
\end{equation}
 Here $|\,.\,|=|\,.\,|_{p^n}=|\,.\,|_p$ is the norm on $\mathbb{Q}_{p^n}$, $k$ is the $p$-adic ``momentum'', and $\widetilde\phi$ is the Fourier transform of $\phi$:
 \be
 \phi(x)=\int\limits_{\mathbb{Q}_{p^n}} dk \chi(k)\widetilde\phi(k x),
 \ee
 where $\chi(x)=\exp{2\pi i \left\{x\right\}}$
 is the additive character on $\cQ_{p^n}$. Dispersion $s$ corresponds to the Vladimirov derivative ``power'' in the configuration space:
 \be
D^s\phi(x)=\frac{1}{\Gamma_p(-s)}\int dy \frac{\phi(y)-\phi(x)}{|y-x|^{1+s}}.
 \ee
Our aim is to compute the one-loop effective potential for the theory given by \eqref{eq:main} with $p, s$ and $n$ fixed. Here $n$ plays the role of space-time ``dimension'' as explained in \cite{Gubser2016}, so one can think of the $n=1$ case as of $p$-adic quantum mechanics, and $n=4$ corresponds to four-dimensional scalar field. Since $\phi$ is real-valued, derivation of the effective potential in general follows the strategy of calculating of the conventional Coleman-Weinberg potential but with a different propagator:
\be
G(k)=\frac{1}{|k|^s}.
\ee
As usual, we split $\phi$ in the background field $\phi_b$ and the dynamical field (see App. \ref{rvCW} for the outline of the conventional calculation), and sum all one-loop diagrams having $2m$ background field external legs each. Assigning $(- \lambda\phi_b^2/2)$ factor to each vertex and taking into account symmetry factor $1/2m$, we write
\be
\Delta \Gamma(\phi_b)= V_{p^n}\underset{m}{\sum}\int  \frac{1}{2m}\Big( \frac{-\lambda \phi_b^2}{2|k|^s} \Big)^m dk=-\frac{V_{p^n}}{2}\int  \ln{\Big(1+\frac{\lambda \phi_b^2}{2|k|^s} \Big)} dk,
\ee
where $V_{p^n}$ is the (infinite) normalization constant corresponding to the volume of ${\mathbb{Q}}_{p^n}$.
Thus the one-loop correction to the effective action is given by
\be
\Delta V = - \frac{\Delta \Gamma(\phi_b)}{V_{p^n}}=\frac{1}{2} \int\limits_{\cQ_{p^n}} dk  \log(1+\frac{\lambda \phi_b^2}{2|k|^s}). \label{eq:dGamma_starting}
\ee
This is a direct analogue of the conventional expression for the Coleman-Weinberg potential.
We should make a remark that here we used $(- \lambda\phi_b^2/2)$ vertex factor from the very beginning instead of taking $ i\lambda\phi_b^2/2$ and performing analytical continuation to Euclidean signature later on. The reason for doing this is that, in the non-Archimedean case, the notion of space-time signature is not well-defined, and the Wick rotation cannot be performed to eliminate the logarithmic singularity and imaginary contributions at small $|k|$. Thus we mimic the Euclidean signature by using a prescription for the vertex that corresponds to real measure in the path integral of the theory.

Since the integrand in \eqref{eq:dGamma_starting} depends only on the ${\mathbb Q}_{p^n}$ norm, one can use the formula (see \eqref{eq:appendix_basic_integral}):
\begin{equation}\label{eq:Qp_integration}
    \int\limits_{{\mathbb Q}_{p^{n}}}f(|x|)dx=(1-p^{-n})\sum^{\infty}_{i=-\infty}p^{ni} f(p^i),
\end{equation}
which leads to the formal expression for one-loop correction to the effective potential:
\begin{equation}
     \Delta V (\phi_b)= \frac12 (1-p^{-n})\sum^{\infty}_{i=-\infty}p^{ni}\ln(1+\frac{\lambda\phi_b^2}{2p^{si}})
    \label{eq:potential_Qp^n}.
\end{equation}
This series is the starting point of our analysis. In a general case, it is divergent, so we have to regularize it by imposing a finite-scale cut-off by analogy with the real case:
\begin{equation}
 \Delta V(\alpha)=\frac12(1-p^{-n})\sum^{M}_{i=-\infty}p^{n i}\ln(1+\alpha p^{-s i})
    \label{eq:potential_Qp^n},
\end{equation}
where $\alpha=\lambda\phi^2_b/2$. One can think of number $M$ as of a logarithm of the corresponding ultraviolet momentum scale $|k|_{UV}=\Lambda=p^M$.

A more subtle feature of $\Delta V$ is that, as a function of background field $\phi_b$, it contains a set of logarithmic singularities at $\phi_b^2=2p^{si}/\lambda$ points if $\lambda<0$.
These points are concentrated around $\phi_b=0$ and $\phi_b=+\infty$. Since we deal with finite $\phi_b$, only the divergences around $\phi_b=0$ matter. 
These singularities arise only for tachyonic expression $\lambda<0$, while the stable  case $\lambda>0$ leads to non-singular expression.

To proceed further, we split sum \eqref{eq:potential_Qp^n} into two parts. For that, we introduce index $I_{\alpha}$ as
\be
I_{\alpha}=[\ln|\alpha|/s\ln p],
\ee
where the brackets denote integer part.
If $i>I_{\alpha}$ and $|\alpha| p^{-si}<1$, logarithm in \eqref{eq:potential_Qp^n} can be expanded as a convergent Taylor series, $\ln(1+ x)= x-x^2/2-\ldots$
For $i<I_{\alpha}$, we rewrite and expand the logarithm as $\ln(1+ x)=\ln( x)+\ln(1+ 1/x)=\ln( x)+ 1/x-1/(2x^2)-\ldots$

Summing these two parts after expansion we obtain an expression that is valid everywhere except for the aforementioned singular points $\alpha = p^{is}$:
\begin{gather}\label{eq:firstest}
    2\Delta V(\alpha)=-(1-p^{-n})\sum^{M}_{i=I_{\alpha}}p^{n i}\sum^{\infty}_{l=1}\frac{(-\alpha)^l p^{-sli}}{l}+(1-p^{-n})\sum^{I_{\alpha}-1}_{i=-\infty}p^{n i}\ln(\alpha p^{-s i})-\\
    -(1-p^{-n})\sum^{I_{\alpha}-1}_{i=-\infty}p^{n i}\sum^{\infty}_{l=1}\frac{p^{sli}}{l(-\alpha)^l }= \nonumber \\ \label{eq:double_expansion}
    -(1-p^{-n})\sum^{\infty}_{l=1}\frac{(-\alpha)^l (p^{(M+1)(n-sl)}-p^{I_{\alpha}(n-sl)})}{l(p^{n-sl}-1)}+(1-p^{-n})\left[- \frac{p^{n(I_{\alpha}-1)}}{p^{-n}-1}\ln(\alpha) -\right.\\ \left.
    -p^{n(I_{\alpha}-2)}\frac{(I_{\alpha}-1)p^{n}-I_{\alpha}}{(1-p^{-n})^2}s\ln p \right]
    -(1-p^{-n})\sum^{\infty}_{l=1}\frac{ p^{I_{\alpha}(n+sl)}}{l(-\alpha)^l(p^{n+sl}-1)}\nonumber
\end{gather}
If $\alpha>0$, the series converges and no issues arise.
If $\alpha<0$, the series in the first line of \eqref{eq:double_expansion} diverges, -- this will be cured by the renormalization procedure.

This sum can be approximated by neglecting integer part operation in $I_\alpha$ and taking $I_{\alpha}=\ln|\alpha|/s\ln p$, so that $p^{I_{\alpha}}=|\alpha|^{\frac{1}{s}}$. We obtain:
\begin{gather}
    2\Delta V(\alpha)=-(1-p^{-n})\sum^{l_0}_{l=1}\frac{(-\alpha)^l \Lambda^{n-sl}}{l(1-p^{sl-n})}+\label{eq:dGamma_no_log}\\+ p^{-n}|\alpha|^{\frac{n}{s}}\ln(\sign\alpha)
     +\frac{p^{-n}}{1-p^{-n}}|\alpha|^{\frac{n}{s}}s\ln p+\nonumber \\
    +(1-p^{-n})|\alpha|^{\frac{n}{s}}\sum^{\infty}_{l=1}\frac{(-\sign\alpha)^l}{l}\Big( \frac{1}{p^{n-sl}-1}-\frac{1}{p^{n+sl}-1}\Big). \nonumber
\end{gather}
Here we introduced $l_0=[n/s]$ to separate the terms that diverge as $\Lambda\rightarrow\infty$. If $l_0=n/s$, a logarithmic term emerges:
\begin{gather}\label{eq:dGamma_log}
    2\Delta V(\alpha)=-(1-p^{-n})\sum^{l_0-1}_{l=1}\frac{(-\alpha)^l \Lambda^{n-sl}}{l(1-p^{sl-n})}-(1-p^{-n})\frac{(-\alpha)^{\frac{n}{s}}}{l_0 s\ln p}\ln\frac{\Lambda^s}{-\alpha}+\\+ p^{-n}|\alpha|^{\frac{n}{s}}\ln(\sign\alpha)
     +\frac{p^{-n}}{1-p^{-n}}|\alpha|^{\frac{n}{s}}s\ln p+\nonumber \\
    +(1-p^{-n})|\alpha|^{\frac{n}{s}}\sum^{\infty}_{l=1, l\neq l_0}\frac{(-\sign\alpha)^l}{l} \frac{1}{p^{n-sl}-1}-(1-p^{-n})|\alpha|^{\frac{n}{s}}\sum^{\infty}_{l=1}\frac{(-\sign\alpha)^l}{l}\frac{1}{p^{n+sl}-1}. \nonumber
\end{gather}
The singular terms can be removed by means of the standard renormalization protocol. For $\alpha>0$, we need to take care only of the terms dependent on $\Lambda$. For  $\alpha<0$, we also need to remove the divergent series (the last sum in \eqref{eq:dGamma_log}).
In both cases, the renormalization conditions are\footnote{Note, that in the case $n=2, s=2$ the renormalization conditions are slightly different, see Sec. \ref{sec:n2s2}}:
\begin{gather} \label{eq:renormalization_conditions}
V^{(4)}_{\phi_b}(\phi_0)=\lambda,\\
V^{''}_{\phi_b}(0)=m_R^2=0,\nonumber
\end{gather}
where we introduced additional scale $\phi_0$ to step away from the logarithmic singularity.

Let us now perform the renormalization procedure for three concrete choices of $n$ and $s$.

\subsection{Case n=1 s=2}
Series \eqref{eq:dGamma_no_log} converges for positive $\lambda$ and contains no terms dependent on $\Lambda$ since $l_0=0$, so we can readily evaluate it without the need to renormalize:
\begin{gather} \label{eq:case_n1_s2}
\Delta V=\frac{1}{2}\sqrt{\frac{|\lambda|}{2}}|\phi_b|\left[ 
     \frac{p^{-1}}{1-p^{-1}}2\ln p+ \right. \left.
     (1-p^{-1})\sum^{\infty}_{l=1}\frac{(-1)^l}{l}\left( \frac{1}{p^{1-2l}-1}-\frac{1}{p^{1+2l}-1}\right) \right]
\end{gather}
Convergence of the latter sum follows from the Leibniz criterion.

If $\lambda<0$, we add a counterterm $A|\phi_b|$ to \eqref{eq:dGamma_no_log} and impose conditions\footnote{Although two renormalization conditions for one counterterm seem like an overdefined problem, they can be consistently resolved.} \eqref{eq:renormalization_conditions}. Then the counterterm exactly cancels the bare terms, and $\Delta V$ becomes trivial. This makes the cases of $\lambda>0$ and $\lambda<0$ qualitatively different. If $\lambda>0$, a term $\sim |\phi_b|$ adds to the effective potential, while for $\lambda<0$ the effective potential does not receive any one-loop corrections.

\subsection{Case n=2 s=2}\label{sec:n2s2}
To perform renormalization, we need to modify conditions \eqref{eq:renormalization_conditions} by shifting the mass renormalization condition to scale $\phi_0$ as well
\be \label{n2s2}
V^{''}_{\phi_b}(\phi_0)=0
\ee
 Solving then equations on $A\phi_b^2$ and $B\phi_b^4$ counterterms, we arrive at:
\begin{equation} \label{eq:case_n2_s2}
    \Delta V = -\frac{\lambda \phi_b^4}{4t^2}+\frac{(1-p^{-2})\lambda\phi_b^2}{2\log p}\left(-1 +\frac{t^2}{24} +\frac{\log t^2 }{4}\right),\,\,t=\phi_b/\phi_0
\end{equation}
This flow is the most non-trivial one among considered cases. Depending on the dimensionless parameter $t$, the effective potential can take different forms with renormalized coupling constant $\lambda_R = \lambda \left(1-6/t^2\right)$ and mass acquiring both positive and negative values. 
\subsection{Case n=4 s=2}
Adding counterterms $A\phi_b^2$ and $B\phi_b^4$ and solving \eqref{eq:renormalization_conditions}, we obtain renormalized one-loop correction to the effective potential of the following form:
\begin{equation}\label{eq:case_n4_s2}
    \Delta V=    \frac{\lambda^2\phi_b^4}{32\log p} \left(1-\frac{1}{p^4}\right)\left(\log \frac{\phi_b^2}{\phi_0^2}-\frac{25}{6}\right).\end{equation}
\vskip20pt
\noindent
Interesting to note that the $p$-adic one-loop corrections to the effective potential have structure very similar to that of their ${\mathbb R}^n$ cousins. Moreover, as we will show in the next section, the Archimedean case can be reproduced from the non-Archimedean one in the formal limit of $p\rightarrow 1$.

\section{$p\to1$ limit}\label{sec:pto1}
One of important reasons why physical theories defined over $p$-adic number fields attract attention is their possible connections to real-domain theories. There are different ways to relate Archimedean and non-Archimedean physical models. The most canonical approach is via adelic formulas, when observables in real theory are decomposed into products over their $p$-adic analogues at all possible values of $p$ \cite{Freund:1987ck,Arefeva:1988kr}. Recently, a construction employing Berkovich spaces was suggested to relate energy spectra of $p$-adic and real quantum mechanics \cite{Huang:2020vwx}. Not widely discussed but elegant approach is based onto $p\rightarrow 1$ limit \cite{Spokoiny:1988zk,Gerasimov:2000zp,Bocardo-Gaspar:2017atv}. To proceed along this line, one first obtains an explicit $p$-dependent expression (e.g., some observable) in non-Archimedean theory and then takes the formal limit $p\to 1$ treating $p$ as a real number.
The approach was taken in \cite{Gerasimov:2000zp} to relate $p$-adic string theory to conventional string field theory.

To rigorously justify this limit, or to even explain why it provides a connection to real space theories, might require quite some effort \cite{Bocardo-Gaspar:2017atv}. However, in our particular case the reason why this limit could lead to meaningful results is rather transparent. Quantum corrections to the effective potential in $p$-adic field theory are given by integrals of the form \eqref{eq:Qp_integration}. For that kind of expression, the following identity can be proven\footnote{To the best of our knowledge, for $n=1$ it was first derived in \cite{Spokoiny:1988zk}.} (see App. \ref{sec:Apppto1}):
\begin{gather}\label{eq:lim_pto1}
\lim_{p\to1}\int_{\mathbb{Q}_{p^n}}f(|x|)dx=\frac{n\Gamma(n/2)}{2\pi^{n/2}}\cdot\int_{\mathbb{R}^n}f(|x|)dx,\,\,\,n>1, \\
\lim_{p\to1}\int_{\mathbb{Q}_{p}}f(|x|)dx=\int_{\mathbb{R}}f(|x|)dx,\,\,\,n=1,
\end{gather}
where r.h.s. integral is exactly what defines corrections to the effective potential in real space field theory modulo the overall volume factor, see e.g. \eqref{eq:real_correction_integral}. This gives us an exact relation between the effective potentials of $p$-adic and real field theories for arbitrary $s$ and $n$.

To illustrate this statement, we shall go through the three particular cases.
First of all, let us make a comment on the validity of Eqs. \eqref{eq:case_n1_s2}-\eqref{eq:case_n4_s2}. Those were derived from \eqref{eq:double_expansion} under assumption that $[\ln |\alpha| / s \ln p] \simeq \ln |\alpha| / s \ln p$ which is valid at the points $|\alpha|p^{-si}=1, \,\, i \in {\mathbb Z}$. In the limit $p\rightarrow 1$, such points form a dense set, and approximation \eqref{eq:dGamma_no_log}-\eqref{eq:dGamma_log} becomes exact, which means that we can  just take $p\rightarrow 1$ limit of \eqref{eq:case_n1_s2}-\eqref{eq:case_n4_s2}.

From that we readily obtain
\begin{itemize}
    \item for $n=1, s=2$:
    \begin{gather}
    \Delta V^{(1,2)}_{p\rightarrow 1} = \frac{1}{2}\theta(\alpha) \lim\limits_{p\rightarrow 1}\left[\frac{p^{-1}}{1-p^{-1}}|\alpha|^{\frac{1}{2}}2\ln p
    + \right. \\ \left. (1-p^{-1})|\alpha|^{\frac{1}{2}}\sum^{\infty}_{l=1}\frac{(-1)^l}{l}\left( \frac{1}{p^{1-2l}-1}-\frac{1}{p^{1+2l}-1}\right) \right] = \nonumber \\
    \left(2 |\alpha|^{\frac12}+
    |\alpha|^{\frac12}\sum^{\infty}_{l=1}\frac{4(-1)^l}{1-4l^2}\right)\theta(\alpha) = \frac{1}{2}\pi |\alpha|^{\frac12}\theta(\alpha)=\frac{1}{2}\pi \sqrt{\frac{|\lambda|}{2}}|\phi_b| \theta(\lambda),\nonumber
\end{gather}
where we introduced Heaviside $\theta$-function to highlight that the one-loop correction is trivial at $\lambda <0$.
\item For $n=2, s=2$:
\begin{equation}
    \Delta V^{(2,2)}_{p\rightarrow 1} = -\frac{\lambda \phi_b^4}{4t^2}+\lambda\phi_b^2\left(-1 +\frac{t^2}{24} +\frac{\log t^2 }{4}\right),\,\,t=\phi_b/\phi_0.
\end{equation}
\item For $n=4, s=2$:
\begin{equation}
    \Delta V^{(4,2)}_{p\rightarrow 1}=
    \frac{\lambda^2\phi_b^4}{8}\left(\log \frac{\phi_b^2}{\phi_0^2}-\frac{25}{6}\right).
\end{equation}
\end{itemize}
Computing the corresponding one-loop corrections in real space field theory, for $n=1$ and $n=4$ we conclude:
\begin{gather}
    \Delta V^{(1,2)}_{\mathbb R} =\frac{1}{4}\sqrt{\frac{|\lambda|}{2}}|\phi_b|= \frac{1}{2\pi}\Delta V^{(1,2)}_{p\rightarrow 1},\\
        \Delta V^{(4,2)}_{\mathbb R} =\frac{\lambda^2\phi_b^4}{256\pi^2}\left(\log \frac{\phi_b^2}{\phi_0^2}-\frac{25}{6}\right)= \frac{1}{32\pi^2}\Delta V^{(4,2)}_{p\rightarrow 1}=\frac{2\pi^2}{(2\pi)^4 4 \Gamma(2)} \Delta V^{(4,2)}_{p\rightarrow 1}. \nonumber
\end{gather}
This looks like a nice evidence supporting our claim. At the same time, the $n=2$ case is more subtle. Before renormalization, the real Coleman-Weinberg potentials perfectly matches its p-adic cousin obtained by means of {\it the Euler-Maclaurin integral approximation} (see Sec. \ref{sec:EM_est} for details, and Eq. \eqref{eq:EM_n2s2_nonrenorm} in particular):
\begin{equation}
    \Delta \widetilde{V}^{(2,2)}_{\mathbb R} = -\frac{\lambda\phi_b^2}{16\pi}\left(1+\ln(\frac{2\Lambda^2}{\lambda\phi^2_b})-\ln(-1)\right) = \frac{1}{4\pi} \Delta \widetilde{V}^{(2,2)}_{p \rightarrow 1} =\frac{2\pi}{2(2\pi)^2\Gamma(1)} \Delta \widetilde{V}^{(2,2)}_{p \rightarrow 1},
\end{equation}
where we use tilde to stress out that those are potentials before renormalization.
After renormalization a mismatch occurs. The reason is that in two dimensions, we have to impose mass renormalization condition at some scale $\phi_0\neq 0$. This leads us to appearance of $\sim 1/t^2$ term in the renormalized potential which comes with different relative coefficients in $p$-adic and in real field theories:
\begin{gather}
    \Delta V^{(2,2)}_{p\rightarrow 1} = -\frac{\lambda \phi_b^4}{4t^2}+\lambda\phi_b^2\left(-1 +\frac{t^2}{24} +\frac{\log t^2 }{4}\right),\\
    \Delta V^{(2,2)}_{\mathbb R} = -\frac{\lambda \phi_b^4}{4t^2}+\frac{\lambda\phi_b^2}{4\pi}\left(-1 +\frac{t^2}{24} +\frac{\log t^2 }{4}\right). \nonumber
\end{gather}
If we assume $\phi_b\gg\phi_0$, this term becomes negligible, and the matching restores.
\section{$p\rightarrow\infty$ limit}
\label{sec:ptoinfty}
Another limit which is instructive to consider is $p\rightarrow\infty$. It seems to exhibit very different behavior from what one can see for any fixed finite $p$. In this case, $I_{\alpha}=[\ln |\alpha|/s\ln p]=0$, and only the first sum in \eqref{eq:double_expansion} survives:
\begin{gather}
    2\Delta V(\alpha)=-(1-p^{-n})\sum^{\infty}_{l=1}\frac{(-\alpha)^l (p^{(M+1)(n-sl)}-1)}{l(p^{n-sl}-1)}
\end{gather}
As before, introducing $l_0=[n/s]$ to separate UV-divergent terms from the rest, we can write
\begin{equation}
    \sum^{\infty}_{l=1}\frac{(-\alpha)^l}{l}\frac{p^{(M+1)(n-sl)}-1}{p^{n-sl}-1}\simeq \sum^{l=l_0}_{l=1}\frac{(-\alpha)^l}{l}[p^{M(n-sl)}-1]-\ln(1+\alpha).
\end{equation}
Note that in contrast with Sec. \ref{sec:main_section}, here we restrict our considerations to $|\alpha|=|\lambda|\phi_b^2/2 <1$. Restoring $\Lambda=p^M$ notation, for the one-loop correction we obtain:
\begin{equation}\label{eq:DeltaGamma_infp_nonlog}
    2\Delta V(\phi_b)=-\sum_{l=1}^{l=l_0}\frac{(-\lambda \phi^2_b)^l}{l2^l}[\Lambda^{n-sl}-1]
        +\ln(1+\frac{\lambda\phi^2_b}{2}).
\end{equation}
If $n=s l_0$, it rather acquires the form:
\begin{equation}\label{eq:DeltaGamma_infp_log}
    2\Delta V(\phi_b)=-\sum^{l=l_0-1}_{l=1}\frac{(-\lambda \phi^2_b)^l}{l2^l}[\Lambda^{n-sl}-1]-\frac{(-\lambda \phi^2_b)^{l_0}}{ l_0 2^{l_0}}[\frac{\ln \Lambda}{\ln p}-1]
    +\ln(1+\frac{\lambda\phi^2_b}{2}).
\end{equation}
Now we shall consider the three cases of interest discussed before.

If $n=1, s=2$, \eqref{eq:DeltaGamma_infp_nonlog} contains no divergent terms, and in the $p\rightarrow \infty$ the effective potential reduces to
\begin{equation}
    V^{(1,2)}_{\mbox{eff}} = \frac{\lambda}{4!} \phi_b^4 + \frac{1}{2}\ln \left(1+\frac{\lambda\phi^2_b}{2} \right)
\end{equation}

If $n=2, s=2$, there is a logarithmic term we need to renormalize. In that case, we do not need to make a shift to some $\phi_0$ scale, and renormalization \eqref{eq:renormalization_conditions} conditions can be imposed at $\phi_b=0$. Adding $A\phi_b^2$ and $B\phi_b^4$ counterterms, we obtain
\begin{equation}
    V^{(2,2)}_{\mbox{eff}} = \frac{\lambda\phi_b^4}{4!}(1+\frac{3}{2}\lambda)-\frac{\lambda \phi_b^2}{4}+\frac{1}{2}\ln \left(1+\frac{\lambda\phi^2_b}{2} \right).
\end{equation}

If $n=4, s=2$, there are both logarithmic and $\Lambda^2$ terms. However, after renormalization we obtain exactly the same result:
\begin{equation}
    V^{(4,2)}_{\mbox{eff}} = \frac{\lambda\phi_b^4}{4!}(1+\frac{3}{2}\lambda)-\frac{\lambda \phi_b^2}{4}+\frac{1}{2}\ln \left(1+\frac{\lambda\phi^2_b}{2} \right).
\end{equation}
A peculiar feature of the $p\rightarrow\infty$ limit is the logarithmic term in the effective potential for all ``space-time'' dimensions.
It does not have an analogue in the conventional real field theory, but does not lead to any unusual or pathological behavior causing neither symmetry breaking nor singularities in the potential if $\lambda >0$.

\section{Euler-Maclaurin estimate of the effective potential}\label{sec:EM_est}
While we managed to compute the effective potential by evaluating series \eqref{eq:dGamma_no_log}-\eqref{eq:dGamma_log}, relying on the assumption that $[\ln \alpha / s\ln p]\simeq \ln \alpha / s\ln p$, it is instructive to discuss another possible approach to do that. Naively, a sum of that kind can be approximated by a continuous integral:
\begin{equation}
    \sum _{j=-\infty}^{M}f(j)\simeq \int _{-\infty}^{M}f(x)\,dx.
\end{equation}
That would be possible if the Euler--Maclaurin formula for infinitely differentiable functions was valid:
\begin{equation}
\label{eq:EMsumformula_inf}
\sum _{i=m}^{M}f(i)=\int _{m}^{M}f(x)\,dx+{\frac {f(M)+f(m)}{2}}+\sum _{k=1}^{\infty}{\frac {B_{2k}}{(2k)!}}(f^{(2k-1)}(M)-f^{(2k-1)}(m)),
\end{equation}
and the residual term was small enough.

Mildly speaking, applicability of this formula in our case is questionable. However, we can plainly compute the integral estimation and make an attempt to relate the outcome of the evaluation to the previously obtained results.

If $n=1, s=2$, the integral converges as $M\rightarrow \infty$ for $\lambda >0$ (the other case becomes trivial after renormalization), and we get:
\be
\Delta V=\frac12 (1-p^{-1})\int^{+\infty}_{-\infty} p^x \ln(1+\frac{\lambda\phi^2_b}{2p^{2x}})dx=(1-p^{-1})|\phi_b|\frac{\sqrt{|\lambda|}\pi}{2\sqrt{2}\ln p}
\ee
versus the result of series summation \eqref{eq:case_n1_s2}:
\begin{gather}
\Delta V=\frac{\sqrt{|\lambda|}}{2\sqrt{2}}|\phi_b| N(p), \\
     N(p)=\frac{p^{-1}}{1-p^{-1}}2\ln p+  (1-p^{-1})\sum^{\infty}_{l=1}\frac{(-1)^l}{l}\left( \frac{1}{p^{1-2l}-1}-\frac{1}{p^{1+2l}-1}\right). \nonumber
\end{gather}
There is a clear discrepancy between these two expressions for large values of $p$ since:
\begin{gather}
    \lim\limits_{p\rightarrow \infty} N(p) = \ln 2, \nonumber \\
    \lim\limits_{p\rightarrow \infty} \frac{\pi(1-p^{-1})}{\ln p} = 0.\nonumber
\end{gather}
On the other hand, for small $p$ the Euler-Maclaurin estimate has surprisingly good accuracy. For example, for $p=7$:
\begin{gather}
    N(7) \simeq 1.387, \nonumber \\ \frac{\pi(1-7^{-1})}{\ln 7} \simeq 1.384. \nonumber
\end{gather}
If $n=2, s=2$, the integral approximation gives:
\begin{equation}
    \Delta V=-\frac{(1-p^{-2})\lambda\phi_b^2}{8\ln p}\left(1+\ln(\frac{2\Lambda^2}{\lambda\phi^2_b})\right), \label{eq:EM_n2s2_nonrenorm}
\end{equation}
which after renormalization with conditions $V''(\phi_0)=0$, $V^{(4)}(\phi_0)=\lambda$, becomes
\begin{equation} \label{eq:case_n2_s2}
    \Delta V = -\frac{\lambda \phi_b^4}{4t^2}+\frac{(1-p^{-2})\lambda\phi_b^2}{2\log p}\left(-1 +\frac{t^2}{24} +\frac{\log t^2 }{4}\right),\,\,t=\phi_b/\phi_0.
\end{equation}
Finally, for $n=4, s=2$:
\begin{gather}
\Delta V=\frac12 \left( 1-p^{-4}\right) \int^{M}_{-\infty} p^{4x}\ln\left(1+\frac{\lambda \phi^2_b}{2p^{2x}}\right)\,dx =\label{eq:Euler_n4s2}\\
\frac{1-p^{-4}}{8\ln p}\left(\Lambda^4\ln\left(1+\frac{\lambda\phi^2_b}{2\Lambda^2}\right)+\frac{\lambda\phi^2_b}{2}\Lambda^{2}-\frac{\lambda^2\phi^4_b}{4}\ln\left(1+\frac{2\Lambda^2}{\lambda\phi^2_b}\right)\right) \simeq \nonumber \\
\frac{1-p^{-4}}{8\ln p}\left(\lambda \phi_b^2 \Lambda^2+\frac{\lambda \phi_b^2}{2}-\frac{\lambda^2\phi^4_b}{4}\left(\ln\frac{2\Lambda^2}{-\lambda\phi^2_b}\right)\right),\,\,\,\Lambda \rightarrow \infty \nonumber
\end{gather}
where we restored $\Lambda=p^{M}$ notation. Renormalization of \eqref{eq:Euler_n4s2} with conditions \eqref{eq:renormalization_conditions} leads to
\begin{equation}
    \Delta V=\frac{\lambda^2\phi_b^4}{32\log p} \left(1-\frac{1}{p^4}\right)\left(\log \frac{\phi_b^2}{\phi_0^2}-\frac{25}{6}\right),
\end{equation}
Contra to the $n=1$ case, for $n=2$ and $n=4$, there is no difference between the Euler-Maclaurin estimate and the discreet sum.
Technically this happens because the coefficients in front of logarithmically divergent terms $\ln (2\Lambda^2/\lambda \phi_b^2)$ are the same in sum \eqref{eq:dGamma_log} and in the Euler-Maclaurin estimate \eqref{eq:Euler_n4s2}. These terms define the form of renormalized potential, while the terms that do not depend on $\Lambda$ and the divergences of higher order are eliminated by renormalization completely. It can be shown that this kind of perfect matching between the Euler-Maclaurin and series expressions for the effective potential always takes place if $n/s\in {\mathbb N}$.

\section{Summary and discussion}
\label{sec:summary}
We have studied one-loop effective potential in the real-valued scalar field theory over unramified extension $\mathbb Q_{p^n}$ of $p$-adic numbers. Typically, by computing the effective potential one can easily gain information on quantum behavior of field theory, since it provides a transparent representation of such concepts as symmetry breaking and renormalization group flow.

In the conventional textbook case, the Feynman diagrams contributing to the effective potential are usually computed in Euclidean signature. In $p$-adic field theory, the Wick rotation is not well defined, hence we have to change the measure of the path integral, simulating Euclidean behavior of the partition function.

For arbitrary fixed $p$, the effective potential is given by a formal series that can be evaluated approximately. In all studied dimensions ($n=1, 2, 4$), the analytical structure of the potential is very similar to that in Archimedean theory, and all the results regarding vacuum stability in conventional $\lambda\phi^4$ theory hold true in the non-Archimedean case. Moreover, in the $p \rightarrow 1$ limit, the effective potential of real field theory can be exactly reproduced from the $p$-adic one. At first glance, this correspondence seems surprising, given the huge difference between real and $p$-adic geometries, and deserves a more detailed discussion. First, an interesting analogy can be drawn with deformations of quantum mechanics. As shown in \cite{Arefeva1991}, in some cases, $q$-deformation of quantum mechanics is related to $p$-adic quantum mechanics with $p=q^{-1}$. It can be possible that the observed similarity between real and $p$-adic effective potentials indicates that the quantum field theory over $p$-adic field can be interpreted as a deformation of the Archimedean one. At the same time, it might well be that the similarity fades away once higher-loop corrections are taken into account. On the level of one loop, the Coleman-Weinberg potential is given by effectively one-dimensional integral \eqref{eq:Qp_integration_app} that can be matched with its real analogue. If there are more than one momentum running in the loops, the real/$p$-adic correspondence can be destroyed, and $p$-adic theory starts qualitatively deviating from its ${\mathbb R}^n$ cousin. We find this aspect interesting and important to investigate.

Another limit we have considered is $p\to\infty$. In contrast with the finite $p$ and $p\rightarrow 1$ cases, it leads to a totally different vacuum structure than in the real field theory. An unusual logarithmic term emerges that  survives the renormalization procedure. If $\lambda<0$, the potential has a singular minimum at $\phi_b = \sqrt{2}/|\lambda|$. However, there is a high chance that it is an artifact of either one-loop approximation or the fact that this limit is singular (i.e. cannot be smoothly derived from finite-$p$ effective potential expression).

Our study leaves a number of open questions. Hopefully, some of them can be answered by computing the next-order quantum corrections which could shed light on what the main difference between $p$-adic and real field theories is. Apart from that, it seems essential to proceed further along the line of studying objects that can be potentially used to clarify the RG flows structure in $p$-adic theories, where one has to deal with scaling transformations in two different number fields. We hope to address these issues in the future.

\section*{Acknowledgements}
We would like to thank Irina Aref'eva and Igor Volovich for useful discussions.
This work was performed at the Steklov International Mathematical Center and supported by the Ministry of Science and Higher Education of the Russian Federation (agreement no. 075-15-2019-1614) and Dutch Science Foundation NWO/FOM under
Grant No. 16PR1024.

\appendix
\section{Definition of $\mathbb{Q}_{p^n}$ space}
In order to describe higher-dimensional structures in $p$-adic mathematical physics, one has to construct a non-Archimedean analogue of ${\mathbb R}^n$ space. 
A direct way to do that would be to simply take an external product ${\mathbb Q}_p^n$ of $n$ copies of $p$-adic field and equip it with a structure of vector space. In many cases this would be sufficient. However, bearing in mind possible applications to the $AdS/CFT$ correspondence, it is desirable to have a space that admits a natural holographic interpretation. ${\mathbb Q}_p^n$ is not a field {\it per se}, and thus does not possess a structure of the Bruhat-Tits tree which would play a role of dual bulk geometry.

This issue can be resolved by using instead of ${\mathbb Q}_p^n$ unramified extension of the p-adic number field ${\mathbb Q}_{p^n}$ of degree $[{\mathbb Q}_{p^n} : {\mathbb Q}_{p}]=n$. 
As a vector space, ${\mathbb Q}_{p^n}$ is isomorphic to ${\mathbb Q}^n_{p}$. To be an unramified extension, it must obey the following requirement. If $L$ and $K$ are two fields, and $L$ is an extension of $K$, we can consider quotients of these fields by their maximal ideals $\ell=L/m_{L}$, $k=K/m_{K}$. Then $k$ is a field extension of $\ell$, and if its' degree is equal to the degree of $L$, so that $[\ell : k]=[L:K]$, $L$ is an unramified extension. Explicitly, ${\mathbb Q}_{p^n}$ can be obtained from ${\mathbb Q}_{p}$ by adjoining a primitive $(p^n - 1)$-st root of unity \cite{Gouvea_textbook}.

We also need to equip ${\mathbb Q}_{p^n}$ with a norm that satisfies the requirement of ultrametricity and becomes the standard $p$-adic norm for $n=1$. It is also handy to assume that the norm takes values in integer powers of $p$, since it induces a branching structure that can serve as a skeleton of the Bruhat-Tits tree. The natural choice is:
\begin{equation}
    |x|=|N(x)|^{1/n}_p, \label{eq:Qpnnorm}
\end{equation}
where $N(x)$ is a determinant of a linear map induced by multiplication in ${\mathbb Q}_{p^n}$: $f(a)=xa$, $a\in {\mathbb Q}_{p^n}$, that can be seen as a linear operator acting on ${\mathbb Q}^n_{p}$.

Integration over ${\mathbb Q}_{p^n}$ is defined in the following way.
As demonstrated in \cite{GubserON}, integral over a constant norm shell in ${\mathbb Q}_{p^n}$ is
\begin{equation}
    \int\limits_{|x|
    =\Lambda}dx=\frac{\Lambda^n}{\zeta(n)},\,\,\,\,\zeta(n)=\frac{1}{1-p^{-n}}.
\end{equation}
Then for a function that depends only on the norm of $p$-adic argument, $f(x)=f(|x|)$:
\begin{equation}
    \int\limits_{{\mathbb Q}_{p^n}} f(|x|)dx=\sum\limits_{\Lambda}\int\limits_{|x|=\Lambda}f(\Lambda)dx=\sum\limits_{\Lambda}f(\Lambda) \Lambda^n \left(1-p^{-n}\right)=\left(1-p^{-n}\right)\sum\limits_{j=-\infty}^{\infty}p^{jn}f(p^j), \label{eq:appendix_basic_integral}
\end{equation}
given $|p^j \frac{m}{q}|=p^{-j}$ (everywhere here norm \eqref{eq:Qpnnorm} is assumed).
\section{The $p\rightarrow 1$ limit in the integral formula}\label{sec:Apppto1}
Here we show that for a general function that depends on the norm of its argument $f(|x|)$, the following identity holds true:
\be\label{eq:Applim_pto1}
\lim_{p\to1}\int_{\mathbb{Q}_{p^n}}f(|x|)dx=\frac{n}{\Omega_n}\cdot\int_{\mathbb{R}^n}f(|s|)d s,
\ee
where $\Omega_n$ is the surface area of an $n$-dimensional unit sphere in real space. The norms on the l.h.s and the r.h.s of \eqref{eq:Applim_pto1} are taken with respect to ${\mathbb Q}_{p^n}$ and ${\mathbb R}^n$ correspondingly.
As we have shown in the previous section:
\begin{equation}\label{eq:Qp_integration_app}
    \int_{{\mathbb Q}_{p^{n}}}f(|x|)dx=(1-p^{-n})\sum^{\infty}_{j=-\infty}p^{nj} f(p^j).
\end{equation}
The sum above is very similar to notion of Jackson integral in $q$-analysis, which is a $q$-deformation of integral over reals and becomes usual integral in the limit of $q\to1$ \cite{KacQuantumCalculus}.

Having a series like \eqref{eq:Qp_integration_app} at hand, one can view $p$ as a formal parameter and continue it from the set of prime numbers onto reals. This makes possible to treat $p$ as a continuous variable and define the $p\to1$ limit. This limit leads to uncertainty in \eqref{eq:Qp_integration_app}: volume element $(1-p^{-n})$ goes to zero and the series itself becomes divergent as an infinite sum of identical constants $f(1)$. To resolve it, we shall rewrite the series as a Darboux sum. We set $p=1+\delta p$, so that $(1-p^{-n})\simeq n\delta p$. The sum \eqref{eq:Qp_integration_app} would then be a Darboux sum for some integral, if the summand function were defined on an equidistant lattice. Rewriting $p^j$ as
\be
p^{j}=(1+\delta p)^{j}=e^{j\ln(1+\delta p)}=e^{j\delta p}, \nonumber
\ee
we obtain:
\be
\lim_{p\to1}(1-p^{-n})\sum^{\infty}_{j=-\infty}p^{nj} f(p^j)= \lim_{\delta p\to0}[n \sum^{\infty}_{j=-\infty}e^{nj\delta p} f(e^{j\delta p})\delta p +o(\delta p)]=n\int^{+\infty}_{-\infty}e^{nx}f(e^{x})dx, \nonumber
\ee
where the summation goes over equally spaced points $j\delta p$, which makes the limit to continuous variable $x=j\delta p$ possible. After a change of variables $k=e^x$, we arrive at \eqref{eq:Applim_pto1}.

Let us illustrate this formula with a simple example of $f(k)=k^s$. Then the followng integral
\be
\int^{1}_{0} k^{n-1}k^sdk=\frac{1}{n+s}, \nonumber
\ee
corresponds to the sum
\be
(1-p^{-n})\sum^{j=0}_{j=-\infty}p^{nj}p^{sj}=\frac{1-p^{-n}}{1-p^{-(n+s)}}, \nonumber
\ee
which in the $p \to 1$ limit gives
\be
\lim_{p\to 1}(1-p^{-n})\sum^{j=0}_{j=-\infty}p^{nj}p^{sj}=\lim_{\delta p\to 0}\frac{1-(1+\delta p)^{-n}}{1-(1+\delta p)^{-(n+s)}}=\frac{n}{n+s}, \nonumber
\ee
and
\be
\lim_{p\to 1}(1-p^{-n})\sum^{j=0}_{j=-\infty}p^{nj}p^{sj}=n\int^{1}_{0} k^{n-1}k^sdk,
\ee
which is a particular case of \eqref{eq:Applim_pto1}.

\section{Coleman-Weinberg potential in conventional scalar field theory}\label{rvCW}
Here we give a short review of the conventional calculation of one-loop Coleman-Weinberg potential in massless scalar field theory:
\be
S=-\int  \left(\phi(x)\partial_\mu \partial^\mu \phi(x) +V(\phi(x))\right) d^nx,\,\,\,\,V(\phi(x))=\frac{\lambda}{4!}\phi^4(x),\,\,\,\, x\in \mathbb{R}^n.
\ee
The prescription is to expand the field $\varphi(x)$ near some stationary background configuration $\phi_b(x)$ as
\be
\phi(x)=\phi_b(x)+\varphi(x) \nonumber
\ee
where $\varphi(x)$ is dynamical field fluctuations, and then derive the quantum corrected effective action for $\phi_b$ by integrating out the dynamical fluctuations. To do that, we expand the action in the path integral up to the second order in $\varphi$:
\begin{gather}
\exp({i \Gamma[\phi_b]})=C(\phi_b) \int D\varphi  \exp \left(i \int d^nx(-\frac{1}{2}\phi\Box\phi-\frac{1}{2}\varphi^2V''(\phi_b)\right), \\
C(\phi_b)=\exp\left({i\int d^nx\left(-\frac{1}{2}\phi_b\Box\phi_b-V(\phi_b)  \right)}\right), \nonumber
\end{gather}
where $\Gamma[\phi_b]$ is the effective action.
The path integral can be evaluated by summing up all one-loop diagrams carrying different number of $\phi_b$ external legs. Each one-loop diagram with $2j$ background field legs comes with a prefactor $(-\frac{i}{2} \lambda \phi_b^2)^j$ and $j$ propagators carrying the same momentum. Thus one has to sum a series of contributions $\mathcal{S}_j$ that have the form
\begin{equation}
    {\mathcal S}_j=\frac{1}{2j}\int\frac{d^n k}{(2\pi)^n}\Big(\frac{\alpha}{k^2+i\varepsilon}\Big)^j, \nonumber
\end{equation}
where we introduced shorthand notation $\alpha =\lambda \phi_b^2 /2$, and the $1/(2j) $ factor is included to account for symmetry of the diagram. Summation of all diagrams gives
\be \label{eq:real_correction_integral}
i\Delta \Gamma= V T\int \frac{d^nk}{(2\pi)^n}\underset{j=1}{\sum} \frac{1}{2n}\Big( \frac{\alpha}{k^2+i\varepsilon}\Big)^j=-VT\frac{1}{2}\int \frac{d^nk}{(2\pi)^n} \ln{\Big(1-\frac{\alpha}{(k^2+i\varepsilon)} \Big)}
\ee
where $V$ and $T$ are divergent normalization constants resulting from volume and time integration. The effective potential is then $\Delta V = -\Delta \Gamma /VT$. Performing the Wick rotation and introducing UV cut-off $\Lambda$, we get
\be
\Delta V= \frac{1}{2}\frac{\Omega}{(2\pi)^{n}}\int_0^\Lambda dk_E k_E^{(n-1)} \log(1+\frac{\alpha}{k_E^2}),
    \ee
    with $\Omega = \Omega_n =2\pi^{\frac{n}{2}}/\Gamma(\frac{n}{2})$ for $n>1$, and $\Omega =1$ for $n=1$.

If $n=1$, the integral converges as $\Lambda \rightarrow \infty $, and
\be
\Delta V=\frac{\sqrt{\lambda} |\phi_b|}{4\sqrt{2}}.
\ee
For $n=4$, after renormalization with \eqref{eq:renormalization_conditions} we arrive at
\be
\Delta V=\frac{\lambda^2\phi_b^4}{256\pi^2}\left(\log\frac{\phi_b^2}{\phi_0^2}-\frac{25}{6}\right).
\ee
In the case of $n=2$, one has to slightly modify the renormalization conditions and use \eqref{n2s2}, which leads to
\be
\Delta V=-\frac{\lambda \phi_b^4}{4t^2}+\frac{\lambda\phi_b^2}{4\pi}\left(-1 +\frac{t^2}{24} +\frac{\log t^2 }{4}\right),\,\,\,\,\,\,t=\phi_b/\phi_0
\ee

\end{document}